\documentclass[12pt]{article}

\usepackage{amsfonts,amssymb,amsmath,latexsym}

\newenvironment{prf}{\begin{trivlist}\item[]{\bf Proof:} }
{\hfill $\Box$ \end{trivlist}}

\newtheorem{thm}{Theorem}
\newtheorem{definition}{Definition}
\newtheorem{prp}[thm]{Proposition}
\newtheorem{lem}[thm]{Lemma}
\newtheorem{cor}[thm]{Corollary}

\newcommand{\bul}{\bullet}

\newcommand{\ep}{{\epsilon}}
\newcommand{\hep}{{\hat\epsilon}}

\newcommand{\RR}{{\mathbb R}}
\newcommand{\CC}{{\mathbb C}}
\newcommand{\HH}{{\mathbb H}}
\renewcommand{\th}{{\theta}}
\newcommand{\sm}{{\sigma}}

\newcommand{\Om}{\Omega}
\newcommand{\tOm}{\tilde\Om}

\renewcommand{\dh}{d_H}
\newcommand{\J}{{\mathcal J}}
\newcommand{\I}{{\mathcal I}}
\newcommand{\G}{{\mathcal G}}
\newcommand{\M}{{\mathcal M}}
\newcommand{\hM}{{\hat{\mathcal M}}}
\newcommand{\cH}{{\mathcal H}}

\newcommand{\cT}{{\mathcal T}}
\newcommand{\cO}{{\mathcal O}}
\newcommand{\al}{{\alpha}}
\newcommand{\be}{{\beta}}
\newcommand{\la}{{\lambda}}
\newcommand{\om}{{\omega}}
\newcommand{\g}{{\mathfrak g}}
\newcommand{\h}{{\mathfrak h}}
\newcommand{\hg}{{\hat{\mathfrak g}}}
\newcommand{\Def}{{\rm Def}}
\newcommand{\no}{\nonumber}
\newcommand{\id}{{\rm id}}
\renewcommand{\part}{\partial}

\newcommand{\op}{\oplus}

\newcommand{\bpart}{{\bar\partial}}

\newcommand{\ot}{\otimes}
\newcommand{\hotimes}{\hat\ot}

\newcommand{\ka}{K\"ahler }
\newcommand{\End}{{\rm End\,}}

\title{On Deformations of Generalized Complex Structures: the Generalized Calabi-Yau Case}

\author{Yi Li\footnote{Current address: \it New High Energy Theory Center, Rutgers University, Piscataway, NJ 08854.}\\
{\small \it California Institute of
Technology, Pasadena, CA 91125, U.S.A.}}

\begin{document}

\begin{titlepage}
\maketitle

\begin{abstract}

We prove an analog of the Tian-Todorov theorem for twisted generalized Calabi-Yau manifolds; namely, we show that the moduli space of generalized complex structures on a compact twisted generalized Calabi-Yau manifold is un-obstructed and smooth. We also construct the extended moduli space and study its Frobenius structure. The physical implications are also discussed.

\end{abstract}
 
\vspace{-5.0in}

\parbox{\linewidth}
{\small\hfill \shortstack{CALT-68-2569}}
\vspace{6.5in}
\end{titlepage}

\section{Introduction}

Generalized complex geometry, recently introduced by Hitchin \cite{Hitchin} and developed more thoroughly by Gualtieri \cite{Gua1}, has attracted growing attention in both mathematics and physics. Mathematically, perhaps the most striking fact is that generalized complex geometry unifies symplectic geometry and complex geometry; this property may shed new light on the mysteries of mirror symmetry \cite{Ben}. From the physical point of view, generalized complex geometry has already found many applications in string theory; see refs.~\cite{Kap, LMT, KapLi, GMPT, CGJ, Zabzine} for a very selected list of examples.

The deformation theory of generalized complex structures is studied by Gualtieri in Ref.~\cite{Gua1}. As in the case of complex geometry where the inequivalent infinitesimal deformations of the complex structure are given by the first cohomology of the tangent sheaf, infinitesimal deformations of a generalized complex structure are also described by a cohomology group: the second cohomology group of a Lie algebroid that is intrinsically associated with the generalized complex structure. This space of infinitesimal deformations is obstructed in general, and the obstruction lies in the third cohomology of the same Lie algebroid. On a generalized complex manifold with non-vanishing third Lie algebroid cohomology, it is usually not a trivial task to determine whether a given infinitesimal deformation is obstructed or not.

To make further progress, one needs to make additional assumptions about or impose additional structures on the generalized complex manifold under consideration. In this article, we investigate the interesting case of generalized Calabi-Yau manifolds, which are generalized complex manifolds with additional structures and whose precise definition will be recalled in Sec.~\ref{sec:GCY}. Generalized Calabi-Yau manifolds are interesting objects because they appear to be the natural candidates on which mirror symmetry acts; furthermore, string theory models based on compactification on generalized Calabi-Yau manifolds have already generated a lot of interest lately. Understanding the deformation theory of generalized Calabi-Yau manifolds is therefore important for both mathematical and physical reasons.

We first recall some known facts about deformations of complex structure on an ordinary Calabi-Yau manifold. Let $X$ be a complex manifold, and $\cT$ its tangent sheaf. As already mentioned, the infinitesimal deformations of the complex structure are parameterized by the sheaf cohomology group $H^1(\cT)$; furthermore, the obstruction map lives in the second sheaf cohomology $H^2(\cT)$ \cite{KS}. In the case of Calabi-Yau manifolds\footnote{There are several inequivalent  definitions of Calabi-Yau manifolds in the literature. The proper definition in our context is that of a K\"ahler manifold with holomorphically trivial canonical bundle or, equivalently, a Riemannian manifold with holonomy group $H\subseteq SU(n)$. See Sec.~\ref{sec:GCY} for more discussion on this point.}, however, one can make a much stronger statement. It was shown by Tian \cite{Tian} and Todorov \cite{Todorov} that the obstruction map vanishes for compact Calabi-Yau manifolds, and therefore the moduli space of complex structures is smooth and of complex dimension $\dim H^1(\cT)$. We prove in this article an analog of the Tian-Todorov theorem for (twisted) generalized Calabi-Yau manifolds. Namely, the moduli space of generalized complex structures on a compact (twisted) generalized Calabi-Yau manifold is un-obstructed and smooth, and has the same dimension as the second cohomology group of the canonically associated Lie algebroid. 

There is a further generalization of the above mentioned results. In modern deformation theory, one assigns a functor to a given algebraic structure to describe its deformations. Under favorable conditions, this deformation functor can be represented by a formal moduli space. This is the case for deformations associated with a generalized Calabi-Yau manifold, where the relevant algebraic structure is that of a differential graded Lie algebra\footnote{More precisely, a differential Gerstenhaber algebra.} that is intrinsically associated with the generalized Calabi-Yau structure. We call the moduli space so constructed the extended moduli space, and it includes the moduli space of generalized complex structures as a subspace. In the case of ordinary Calabi-Yau manifold, the extended moduli space was constructed and analyzed by Barannikov and Kontsevich in Ref.~\cite{BaKon}. In this paper, we extend this analysis to twisted generalized Calabi-Yau manifolds, and comment on the physical significance.

This paper is organized as follows. We recall some basic facts of generalized complex geometry and generalized \ka geometry in Sec.~\ref{sec:GC}; this also serves to explain our conventions and notation. Then we briefly review the deformation theory of generalized complex structures in Sec.~\ref{sec:def}. In Sec.~\ref{sec:GCY}, after giving some preliminary definitions about generalized Calabi-Yau manifolds, we proceed to prove a vanishing theorem (Theorem \ref{thm:vanishing}) for compact generalized Calabi-Yau manifolds. This is one of the main results in this paper. In Sec.~\ref{sec:extended}, we construct the extended moduli space for generalized Calabi-Yau manifolds and study its key properties (the Frobenius structure). Finally, we give physical interpretation of our results in Sec.~\ref{sec:physics}.

\section{Generalized Complex Manifolds and Generalized K\"ahler Manifolds}
\label{sec:GC}

In this section, we recall some basic definitions in generalized complex geometry and generalized \ka geometry. The reader is referred to Ref.~\cite{Gua1} for a more comprehensive treatment.

Let $M$ be a smooth manifold, and $T$ its tangent bundle. The hallmark of generalized geometries is that geometric structures are defined on $T\op T^*$, rather than on $T$ or $T^*$ alone \cite{Hitchin}. A noted difference of $T\op T^*$ from $T$ is that it is equipped with a canonical nondegenerate inner product, which is given by
\begin{equation}
\langle X+\xi, Y+\eta\rangle = \frac12(\iota_X\eta + \iota_Y\xi), \quad \forall \; X,Y\in C^\infty(T), \; \xi,\eta\in C^\infty(T^*).\nonumber
\end{equation}
There is also a canonical bracket on $T\op T^*$, the Courant bracket, which is defined by
$$
[X+\xi, Y+\eta] = [X,Y]+L_X\eta - L_Y\xi -\frac12 d(\iota_X\eta-\iota_Y\xi).
$$
Here, on the right-hand side, $[,]$ denotes the usual Lie bracket of vector fields. Equipped with this bracket structure, $T\op T^*$ is an example of the so-called {\rm Courant algebroids} \cite{LWX}. When restricted to an involutive isotropic subbundle of $T\op T^*$, the Courant bracket becomes a Lie bracket, and the Courant algebroid structure reduces to a Lie algebroid structure\footnote{For an in-depth treatment of Lie algebroids, see Ref.~\cite{Mack}.}.

A generalized almost complex structure on $M$ is a smooth section $\J$ of the endomorphism bundle $\End(T\op T^*)$, which is orthogonal with respect to the canonical inner product, and squares to minus identity on $T\op T^*$. Like almost complex structures, generalized almost complex structures only exist on even-dimensional manifolds. Let $E\subset (T\op T^*)\otimes\CC$ be the $+i$-eigenbundle of $\J$. The generalized almost complex structure is said to be integrable if $E$ is closed under the Courant bracket. In this case, $\J$ is called a generalized complex structure. Note that $E$ is a maximal isotropic subbundle of $(T\op T^*)\otimes\CC$, so the Courant algebroid structure of $T\op T^*$ endows $E$ with the structure of a complex Lie algebroid. This association of a  Lie algebroid to a generalized complex structure is important. Certain cohomology of the Lie algebroid governs the infinitesimal deformations of the generalized complex structure $\J$ \cite{Gua1}, as we shall see in Sec.~\ref{sec:def}. 

On a generalized almost complex manifold $(M,\J)$ of real dimension $2N$, there is an alternative grading on the space of smooth complex differential forms:
\begin{equation}
\label{eq:grading}
\Om^\bul(M,\CC)\simeq U_0\op U_1\op\cdots\op U_{2N}.
\end{equation}
Here $U_0$ is the canonical line bundle of $(M,\J)$, whose sections are forms annihilated by $E$ via the spin representation:
$$
(X+\xi)\cdot \al = \iota_X\al + \xi\wedge \al, \quad \forall\, X\in C^\infty(T), \,Y\in C^\infty(T^*),\, \al\in\Om^\bul(M,\CC).
$$
The spin representation induces a natural action of $\wedge^\bul E^*$ by linearity, and $U_k\equiv\wedge^k {E^*}\cdot U_0$. In the following, we adopt the convention that the action of $A = A_1\wedge A_2\wedge\cdots\wedge A_k\in C^\infty(\wedge^k {E^*})$ on $\Om^\bul(M,\CC)$ be given by 
$$
A\cdot\al = (A_1\wedge A_2\wedge\cdots\wedge A_k)\cdot\al \equiv A_1\cdot A_2\ldots A_k\cdot \al, \quad \forall \, \al\in\Om^\bul(M,\CC).
$$
According to this convention, $(A\wedge B)\cdot\al = A\cdot B\cdot \al$ for any $A, B\in C^\infty(\wedge^\bul E^*)$. 

The alternative grading in (\ref{eq:grading}) can be used to define the generalized Dolbeault operators \cite{Gua1}, which we denote by $\part_\J$ and $\bpart_\J$ to signify their dependence on the generalized almost complex structure $\J$. It can be shown that $\J$ is integrable if and only if $d=\part_\J + \bpart_\J$.

A generalized K\"ahler structure is a triple $(M,\I,\J)$, such that $\I$ and $\J$ are a pair of commuting generalized complex structures on $M$, and that $\G: (A,B)\mapsto -\langle \I\J A,B\rangle$ defines a positive-definite metric on $T\op T^*$. An important property of generalized K\"ahler manifolds is that they satisfy a generalized version of the \ka identities \cite{Gua2}. Let $(M,\J_1,\J_2)$ be a compact generalized K\"ahler manifold, and let $\part_\J$, $\bpart_\J$ be the generalized Dolbeault operators for either $\J_1$ or $\J_2$. Let $\Delta_d$, $\Delta_{\part_\J}$, and $\Delta_{\bpart_\J}$ be the canonically associated Laplacians for $d$, $\part_\J$, and $\bpart_\J$, respectively.\footnote{These Laplacians are constructed by using the Born-Infeld metric, which derives from the generalized Riemannian metric $\G$. See \cite{Gua2} for details.} The generalized \ka identities then imply
\begin{equation}
\label{eq:Hodge}
\Delta_d = 2\Delta_{\part_\J} = 2\Delta_{\bpart_\J}.
\end{equation}
A direct consequence of (\ref{eq:Hodge}) is that a generalized version of the $\part\bpart$-lemma holds for $\part_\J$ and $\bpart_\J$, a fact that will be important for us in the following. Aspects of the generalized $\part\bpart$-lemma in the context of generalized geometries are discussed by Cavalcanti \cite{Cava}.

Everything said so far can be twisted by a real closed 3-form $H\in\Om^3_{\rm closed}(M)$. The  twisted Courant bracket is defined by
$$
[X+\xi,Y+\eta]_H = [X+\xi,Y+\eta] +\iota_Y\iota_X H, \quad \forall\; X,Y\in C^\infty(T), \xi,\eta\in C^\infty(T^*).
$$
One defines twisted generalized complex structures and twisted generalized \ka structures in an obvious way, replacing the Courant bracket by its twisted counterpart everywhere. The alternative decomposition of differential forms in (\ref{eq:grading}) and the definition of generalized Dolbeault operators remain essentially unchanged, except that the integrability condition now reads $d_H = \part_{\J,H}+\bpart_{\J,H}$, where $d_H=d-H\wedge$ is the twisted de Rham differential. There is also a twisted version of (\ref{eq:Hodge}) on a compact twisted generalized \ka manifold:
\begin{equation}
\label{eq:Hodge_H}
\Delta_{d_H} = 2\Delta_{\part_H} = 2\Delta_{\bpart_H}.
\end{equation}
Here and in the following, we write $\part_H$ and $\bpart_H$ for $\part_{\J,H}$ and $\bpart_{\J,H}$ when there is no chance for confusion.

\section{Deformation Theory of Generalized Complex Structures}
\label{sec:def}

In this section, we review the deformation theory of generalized complex structures developed by Gualtieri \cite{Gua1}. If one recalls the relation between a generalized complex structure and its associated Lie algebroid $E$, deformations of $\J$ can be regarded as variations of $E$ inside the space of maximal isotropic subbundles of real index zero, with some extra integrability condition. An admissible variation of $E$ is represented by an element $\ep\in C^\infty(\wedge^2E^*)$, which can be regarded as a bundle map $E\to E^*: A \mapsto A\lrcorner \ep$. It deforms the eigenbundle $E$ to $E_\ep = (1+\ep)E$. It turns out that $E_\ep$ corresponds to an integrable deformation of $\J$ if and only if the Maurer-Cartan equation is satisfied:
\begin{equation}
\label{eq:MC}
d_E\ep + \frac12[\ep,\ep] \;=\; 0.
\end{equation}
Here the bracket denotes the Schouten bracket on $\wedge^\bul E^*$, which is derived from the Lie bracket on $E^*$, and $d_E$ is the Lie algebroid differential on $\wedge^\bul E^*$. Infinitesimal deformations, which are solutions to the linearized Maurer-Cartan equation, $d_E\ep=0$, can be shown to correspond to the second Lie algebroid cohomology $H^2(d_E)$ \cite{Gua1}. 

Given an infinitesimal deformation, there are obstructions to integrating it into a finite deformation of the generalized complex structure that satisfies the Maurer-Cartan equation (\ref{eq:MC}). The obstructions can be shown to live in $H^3(d_E)$. More precisely, one has the following theorem. 

\begin{thm}[Gualtieri, \cite{Gua1}] 
\label{thm:Gua}
There exists an open neighborhood $U\subset H^2(d_E)$
containing zero that parameterizes a smooth family of generalized 
almost complex deformations of $\J$, and an analytic obstruction 
map $\Phi:U\rightarrow H^3(d_E)$ with $\Phi(0)=0$ and $d\Phi(0)=0$, 
such that the integrable deformations are the ones in the sub-family 
$\M=\Phi^{-1}(0)$. Any sufficiently small deformation of $\J$ is
equivalent to at least one member of the family $\M$. In the case that 
the obstruction map vanishes, $\M$ is a smooth locally complete family.
\end{thm}

On a generalized complex manifold with $H^3(d_E)=0$, the obstruction map vanishes automatically, and $M$ admits a smooth family of deformations of dimension $\dim H^2(d_E)$ according to Theorem \ref{thm:Gua}. However, the assumption $H^3(d_E)=0$ is neither necessary nor natural. As we show in the next section, there is a far more interesting class of manifolds that also enjoy the special property of having a vanishing obstruction map.

\section{Deformations on a Compact Twisted Generalized Calabi-Yau Manifold}
\label{sec:GCY}

\subsection{Some definitions}
There are several different notions of generalized Calabi-Yau manifolds used in the mathematics and physics literature. To avoid possible confusions, we introduce some definitions to differentiate among them.  We start with the most general one, which is also the original definition introduced by Hitchin.

\begin{definition}[Hitchin, \cite{Hitchin}]
A weak generalized Calabi-Yau manifold is a generalized complex manifold $(M,\J)$, such that its canonical line bundle is generalized holomorphically trivial, i.e., there exists a nowhere zero section $\Om$ of the canonical bundle such that $d\Om=\bpart_\J\Om=0$. If there is a closed 3-form $H$, one defines a weak twisted generalized Calabi-Yau structure similarly, by requiring $d_H\Om=\bpart_H\Om=0$. The section $\Om$ is sometimes referred to as a generalized holomorphic volume form on $(M,\J)$.
\end{definition}

The scope of weak generalized Calabi-Yau manifolds is rather broad; in particular, any symplectic manifold is a weak generalized Calabi-Yau manifold. To make analogy with an ordinary Calabi-Yau manifold, which usually comes with (or admits) a \ka structure, we should impose additional structures. This motivates the following series of increasingly restrictive definitions.
\begin{definition}
A generalized Calabi-Yau manifold is a triple $(M,\J;\I)$, such that it is a generalized \ka manifold with commuting generalized complex structures $(\I,\J)$, and that $(M,\J)$ defines a weak generalized Calabi-Yau structure.
\label{def:GCY}
\end{definition}
\begin{definition}
A strong generalized Calabi-Yau manifold is a generalized Calabi-Yau manifold $(M,\J;\I)$, such that the canonical line bundle for $\I$ is topologically trivial.
\end{definition}
\begin{definition}[Gualtieri, \cite{Gua1}]
A  generalized Calabi-Yau metric is a generalized \ka manifold $(M,\I,\J)$, such that both $(M,\I)$ and $(M,\J)$ define weak generalized Calabi-Yau structures. 
\end{definition}
Obviously, there is also an $H$-twisted version for each of them.

\vspace{3mm}

Where do ordinary Calabi-Yau manifolds fit in? Unfortunately this is an imprecise question, since there are several slightly different but inequivalent definitions of Calabi-Yau manifolds in the literature as well\footnote{See Ref.~\cite{Joyce}, Chap.~5, for a list of different definitions.}. A commonly used version is that of a \ka manifold whose first Chern class is trivial; this is also the type of geometric structure studied in Yau's proof of the Calabi conjecture \cite{Yau}. According to our terminology, it is actually an example of {\em strong} generalized Calabi-Yau manifolds, with the \ka form defining a weak generalized Calabi-Yau structure. Another popular definition of a Calabi-Yau manifold is a Riemannian manifold with $SU(n)$ holonomy, and its popularity largely comes from string compactification applications. In our terminology, it corresponds to a generalized Calabi-Yau metric. 

In this section, we shall prove a vanishing theorem for compact (twisted) generalized Calabi-Yau manifolds (Theorem \ref{thm:vanishing}). It is a generalization of the well-known Tian-Todorov theorem \cite{Tian, Todorov}, which states that the moduli space of complex structures on a compact Calabi-Yau manifold (in the sense of an $SU(n)$ holonomy manifold) is un-obstructed, to the realm of generalized geometries. As pointed out above, an $SU(n)$ manifold is an example of generalized Calabi-Yau metrics, which are quite more restrictive than the generalized Calabi-Yau structure given in Def.~\ref{def:GCY}. Our results then suggest that even in the traditional setting (i.e., in the `non-generalized' setting), the results of Tian and Todorov can be extended beyond $SU(n)$ holonomy manifolds\footnote{In particular, they can be extended to manifolds that do not admit \ka structure.}. Indeed, this fact was already pointed out by Tian in Ref.~\cite{Tian}. Generalized geometries provide a natural framework for this extension. 

But is the generalized Calabi-Yau structure (as in Definition \ref{def:GCY}) a necessary condition for the  vanishing theorem to hold? Strictly speaking, this is not so. As it will become clear in the next subsection, provided that one has a weak generalized Calabi-Yau structure, and that a form of the identity (\ref{eq:Hodge}) holds, the vanishing theorem should still hold. However, we do not know at this point how to categorize the most general manifolds that have these properties in simple terms. In this article, we shall focus exclusively on generalized Calabi-Yau manifolds, even though the results are applicable to more general situations.

\subsection{The vanishing theorem}

First we prove some preliminary results. Let $(M,\J)$ be an $H$-twisted generalized complex manifold, and $E$ the $+i$-eigenbundle of $\J$ as before. Recall that the space of smooth complex differential forms are decomposed into eigenspaces of $\J$ as in Eq.~(\ref{eq:grading}). The twisted Courant bracket restricts to a Lie bracket on $E^*\simeq \bar{E}$. This Lie bracket can be extended into a Schouten bracket on $\wedge^\bul E^*$, which we continue to denote by $[,]_H$; it is defined by
$$
[A,B]_H =\sum_{i,j}(-1)^{i+j}[A_i,B_j]_H\wedge A_1\wedge\cdots\hat{A_i}\cdots\wedge A_p\wedge B_1\wedge\cdots\hat{B_j}\cdots\wedge B_q,
$$
where $A=A_1\wedge\cdots\wedge A_p\in C^\infty(\wedge^p E^*)$ and 
$B=B_1\wedge\cdots\wedge B_q\in C^\infty(\wedge^q E^*)$.

\begin{lem}
\label{lem:identity}
Let $\rho$ be an arbitrary smooth differential form. For any $A\in C^\infty(\wedge^p E^*)$, $B\in C^\infty(\wedge^q E^*)$, one has
\begin {eqnarray}
\label{eq:identity}
\dh(A\cdot B\cdot\rho) &=& (-1)^p A\cdot \dh(B\cdot\rho)
+(-1)^{(p-1)q}B\cdot \dh(A\cdot\rho)\nonumber\\
&& +\; (-1)^{p-1}[A,B]_H\cdot\rho +(-1)^{p+q+1}A\cdot B\cdot d_H\rho
\end{eqnarray}
\end{lem}

\begin{prf} 
It suffices to assume $A=A_1\wedge A_2\wedge\cdots \wedge A_p$, $B=B_1\wedge B_2\wedge\cdots\wedge B_q$, where all $A_i$'s and $B_i$'s are smooth sections of $E^*$. We proceed by induction. The case of $p=q=1$ is already proven in the literature. It is proved in Ref.~\cite{Gua1} that
$$
A\cdot B\cdot d\rho = d(B\cdot A\cdot\rho)+
B\cdot d(A\cdot\rho)-A\cdot d(B\cdot\rho)+[A,B]\cdot\rho - d\langle A,B\rangle\wedge \rho
$$
for any $A,B\in C^\infty(T\op T^*)$. The $H$-twisted version is proved in Ref.~\cite{KapLi}. As we are restricting to $A, B\in C^\infty(E^*)$, the term involving the inner product can be dropped, and we obtain (\ref{eq:identity}) for $p=q=1$.

Now suppose (\ref{eq:identity}) holds for all $p\le k, q=\ell$. We show that it also holds for $p=k+1, q=\ell$. Assume $A=A_0\wedge \tilde{A}$, where $\tilde{A}=A_1\wedge A_2\wedge\cdots \wedge A_k$, and $B=B_1\wedge B_2\wedge\cdots\wedge B_\ell$, with all $A_i, B_i \in C^\infty(E^*)$. Write $A_0=X+\xi$, where $X\in C^\infty(T)$ and $\xi\in C^\infty(T^*)$. We have
\begin{eqnarray}\label{eq1}
\dh(A\cdot B\cdot\rho) &=& \dh(\iota_X+\xi\wedge)(\tilde{A}\cdot B\cdot\rho)\no\\
&=& (L_X+d\xi\wedge-\iota_X H\wedge)\tilde{A}\cdot B\cdot\rho - A_0\cdot \dh(\tilde{A}\cdot B\cdot\rho)\no\\
&=& (L_X+d\xi\wedge-\iota_X H\wedge)\tilde{A}\cdot B\cdot\rho\no\\
&& -\; A_0\cdot\big((-1)^k \tilde{A}\cdot \dh(B\cdot\rho) + (-1)^{(k-1)\ell}B\cdot \dh(\tilde{A}\cdot\rho)\no\\
&& +\;(-1)^{k-1}[\tilde{A},B]_H\cdot\rho + (-1)^{k+\ell+1}\tilde{A}\cdot B\cdot d_H\rho\big)\no\\
&=& (L_X+d\xi\wedge-\iota_X H\wedge)\tilde{A}\cdot B\cdot\rho\no\\
&& +\; (-1)^{k+1}       A\cdot\dh(B\cdot\rho) - (-1)^{k\ell}B\cdot A_0\dh(\tilde{A}\cdot\rho)\no\\
&& +\; (-1)^k A_0\cdot[\tilde{A},B]_H\cdot\rho + (-1)^{k+\ell}A\cdot B\cdot d_H\rho\no\\
&=&  (-1)^{k+1}A\cdot \dh(B\cdot\rho)+(-1)^{k\ell}B\cdot \dh(A\cdot\rho)\no\\
&& +\; (-1)^{k+\ell}A\cdot B\cdot d_H\rho + (-1)^{k} (A_0\wedge[\tilde{A},B]_H)\cdot\rho\no\\
&& +\;  (L_X+d\xi\wedge-\iota_XH\wedge)\tilde{A}\cdot B\cdot\rho \no\\
&& -\;  (-1)^{k\ell} B\cdot(L_X+d\xi\wedge-\iota_XH\wedge)\tilde{A}\cdot\rho
\end{eqnarray}
Furthermore, we note that for any $C=Y+\eta\in C^\infty(E^*)$ and any $\al\in\Om^\bul(M,\CC)$,
\begin{eqnarray}
[L_X+d\xi\wedge-\iota_X H\wedge,C\cdot]\al &=& 
[L_X+d\xi\wedge-\iota_X H\wedge,\iota_Y+\eta\wedge]\al\no\\
&=& \big(\iota_{[X,Y]}+L_X\eta-\iota_Y(d\xi)\wedge+
        \iota_Y\iota_X H\wedge\big)\al\no\\
&=& [A_0,C]_H\cdot\al.\no
\end{eqnarray}
It is then straightforward to check that the last two terms on the right-hand side of (\ref{eq1}) combine to give
$$
(-1)^{k\ell}\,([A_0,B]_H\wedge \tilde{A})\cdot\rho.
$$
Substituting it back to (\ref{eq1}) yields
\begin{eqnarray}
\dh(A\cdot B\cdot\rho) &=& (-1)^{k+1}A\cdot \dh(B\cdot\rho)+
(-1)^{k\ell}B\cdot \dh(A\cdot\rho)\no\\
        && +\; (-1)^k\,[A,B]_H\cdot\rho + (-1)^{k+\ell}A\cdot B\cdot d_H\rho.\no
\end{eqnarray}
This proves (\ref{eq:identity}) for $q=\ell$ and all $p\ge1$. Since (\ref{eq:identity}) is graded symmetric in $A$ and $B$, it must hold for all $p$ and $q$.
\end{prf}

If $(M,\J)$ is an $H$-twisted weak generalized Calabi-Yau manifold, and $\Om$ a nowhere zero $d_H$-closed section of $U_0$ (i.e., a generalized holomorphic volume form), then  any $A\in C^\infty(\wedge^\bul E^*)$ is canonically associated with a differential form $A' \equiv A\cdot\Om$ via the spin representation. One has the following corollary.

\begin{cor} 
\label{cor:identity2}
Let $A\in C^\infty(\wedge^p E^*), B\in C^\infty(\wedge^q E^*)$. If both $A'$ and $B'$ are $\part_H$-closed, then the following identity holds
\begin{equation}
\part_H(A\wedge B)' \;=\; (-1)^{p-1}\,[A,B]_H'.\nonumber
\end{equation}
\end{cor}
\begin{prf}This is an immediate consequence of Lemma \ref{lem:identity} by letting $\rho=\Om\in C^\infty(U_0)$ and projecting Eq.~(\ref{eq:identity}) to $U_{p+q-1}$. 
\end{prf}

In fact, the map $':C^\infty(\wedge^\bul E^*)\to \Om^\bul(M,\CC)$ establishes a complex isomorphism between $(C^\infty(\wedge^\bul E^*),d_E)$ and $(\Om^\bul(M,\CC), \bpart_H)$. This fact follows from the assertion that $(\Om^\bul(M,\CC), \bpart_H)$ is a Lie algebroid module over the differential graded complex $(\wedge^\bul E^*,d_E)$ \cite{Gua1}. Due to the importance of this result in the following, we provide an independent proof here. 

\begin{prp}\label{prp:iso}
For any $\al\in C^\infty(\wedge^\bul E^*)$, we have 
\begin{equation}
(d_E\al)' \;=\; \bpart_H(\al').
\label{eq:comp_iso}
\end{equation}
\end{prp}
\begin{prf}
It suffices to assume $\al$ to have a definite grade, and we prove the proposition by induction. For any $f\in C^\infty(M)$, $A\in C^\infty(E)$, we have
$$
A\cdot(d_E f)' = (A\lrcorner d_E f)\Om = \pi(A) f\Om = A\cdot df\wedge\Om
= A\cdot d_H(f\Om)
$$
where $\pi$ denotes the anchor map of $E$. As $A$ is arbitrary, one has $(d_Ef)\cdot\Om = d_H(f\Om)$. Now suppose $\al\in C^\infty(E^*)$. Given any $A,B\in C^\infty(E)$,
\begin{eqnarray}
A\cdot B\cdot (d_E\al)' &=& A\cdot (B\lrcorner d_E\al)\cdot\Om\no\\
        &=& \,d_E\al (B,A)\Om\no\\
        &=& \big(-\pi(A)\al(B)+\pi(B)\al(A)+\al([A,B])\big)\Om.\no
\end{eqnarray}
On the other hand, if we denote $B=Y+\eta, Y\in C^\infty(T), \eta\in C^\infty(T^*)$, 
\begin{eqnarray}
A\cdot B\cdot\bpart_H(\al') &=& A\cdot B\cdot d_H(\al\cdot\Om)\no\\
        &=& A\cdot(L_Y+d\eta-\iota_YH)\al'-A\cdot d_H (B\cdot\al\cdot\Om)\no\\
        &=& [A,B]_H\cdot\al\cdot\Om+B\cdot d_H (A\cdot\al')-A\cdot d_H (\al(B)\Om)\no\\
        &=& \al([A,B])\Om+B\cdot d_E(\al(A))\cdot\Om- A\cdot d_E(\al(B))\cdot\Om\no\\
        &=& \big(\al([A,B])+\pi(B)\al(A))-\pi(A)\al(B)\big)\Om.\no
\end{eqnarray}
As $A,B$ are arbitrary, the relation (\ref{eq:comp_iso}) holds for $\al\in C^\infty(E^*)$. By the standard induction argument, it is easy to show that (\ref{eq:comp_iso}) is true for any $\al\in C^\infty(\wedge^\bul E^*)$.
\end{prf}
A direct consequence of this proposition is that the $d_E$-cohomology and the $\bpart_H$-cohomology are isomorphic and can be used interchangeably. 

\vspace{3mm}

We are now ready to state the main theorem of this section. 

\begin{thm}
Let $(M,\J;\I)$ be a compact $H$-twisted generalized Calabi-Yau manifold as in Definition \ref{def:GCY}. Then the obstruction map for deformations of $\J$ vanishes. In particular, sufficiently small deformations of $\J$ are parameterized by a smooth moduli space $\M$ of complex dimension
$\dim_\CC H^2(d_E).$
\label{thm:vanishing}
\end{thm}

\begin{prf}
Let us first define several (quasi-)elliptic differential operators. Recall that $\G=-\I\J$ defines a positive-definite metric on $T\op T^*$, and it induces a natural Hermitian inner product on $\Om^\bul(M,\CC)$ (therefore also on $C^\infty(\wedge^\bul E^*)$) that generalizes the familiar Hodge inner product in ordinary Riemannian geometry \cite{Gua2}. Let us denote this Hermitian inner product by $(\al,\be)$ for $\al,\be\in\Om^\bul(M,\CC)$. Let $\bpart_H, \part_H$ be the generalized Dolbeault operators associated with $\J$. Their adjoints with respect to this Hermitian product are denoted by $\bpart_H^\dagger$ and $\part_H^\dagger$, respectively. Let $G$ be the Green operator for the Laplacian $\Delta_{\bpart_H}$, and $Q$ the smoothing operator $Q = \bpart_H^\dagger G.$ We have the following useful identity on $\Om^\bul(M,\CC)$:
\begin{equation}
\label{eq:decomposition}
\id = \HH + \bpart_H Q + Q\bpart_H,
\end{equation}
where $\HH$ denotes the orthogonal projection to the $\bpart_H$-harmonics. Due to the relation (\ref{eq:Hodge}), there is no need to distinguish $\bpart_H$, $\part_H$, or $d_H$-harmonics. We denote the space of harmonic forms by $\cH^\bul(M)$, with the grading specified by (\ref{eq:grading}).

We prove the theorem by showing that for any infinitesimal deformation, which is parameterized by a class $[\th]\in H^2(d_E)$, one can always construct a power series solution to the Maurer-Cartan equation (\ref{eq:MC}) in the form of
$$
\ep(t) = \sum_a \ep_a t_a + \frac12 \sum_{a_1,a_2}\ep_{a_1a_2}t_{a_1}t_{a_2} + \cdots
\; \in C^\infty(\wedge^2 E^*)\ot\CC[[t_\h]],
$$
where $\h=H^2(d_E)$, such that 
\begin{itemize}
\item[1)]{$\ep_a$ are a basis of $\cH^2(M)$ and $\ep_a t_a$ is the harmonic representative of $[\th]$;}
\item[2)]{all $\ep_{a_1\ldots a_k}, k\ge2$, are $\part_H$-exact and $\bpart_H^\dagger$-closed.}
\item[3)]{$\ep(t)$ is uniformly convergent in a neighborhood of $0$.}
\end{itemize}

For simplicity of the notation, we shall prove the following equivalent result: for any class $[\th]\in H^2(d_E)$, there is a power series solution to the Maurer-Cartan equation in the form of
\begin{equation}
\ep = \sum_{i\ge1} \ep_i t^i, \qquad t\in\CC, \quad \ep_i\in C^\infty(\wedge^2E^*), \;\forall\,i,\no
\end{equation}
such that 
\begin{itemize}
\item[1')]{$\ep_1$ is the harmonic representative of $[\th]$;}
\item[2')]{all $\ep_{k}, k\ge2$, are $\part_H$-exact and $\bpart_H^\dagger$-closed;}
\item[3')]{$\ep(t)$ converges uniformly in a neighborhood of $t=0$.}
\end{itemize}

To show this, let us rewrite the Maurer-Cartan equation using (\ref{eq:comp_iso}):
\begin{equation}
\bpart_H\ep' + \frac12[\ep,\ep]' = 0,\nonumber
\end{equation}
which is equivalent to the following series of equations:
\begin{eqnarray}
\label{eq:MC.series}
\bpart_H\ep_1' &=& 0\no\\
\bpart_H\ep_2' &=& - \frac12[\ep_1,\ep_1]'\no\\
&\vdots&\\
\bpart_H\ep_n' &=& - \frac12\sum_{i=1}^{n-1}[\ep_i,\ep_{n-i}]'\no\\
&\vdots&\no
\end{eqnarray}
The first equation in (\ref{eq:MC.series}) is satisfied if we take $\ep_1$ to be the $d_E$-harmonic representative of $[\th]$. This is equivalent to $\ep_1'\in \cH^2(M)$. 

By Corollary \ref{cor:identity2}, $\psi_2\equiv [\ep_1,\ep_1]^{'}$ is $\part_H$-exact. By the general results of Hodge theory, $\psi_2$ has no harmonic component, i.e., $\HH\psi_2=0$. It is also $\bpart_H$-closed, because
\begin{eqnarray}
\bpart_H\psi_2 &=& \left(d_E[\ep_1,\ep_1]\right)'\no\\
        &=& \left([d_E\ep_1,\ep_1]-[\ep_1,d_E\ep_1]\right)'\no\\
        &=& 0.\nonumber
\end{eqnarray}
Applying the identity (\ref{eq:decomposition}) on $\psi_2$, we obtain $\psi_2 = \bpart_H Q\psi_2$. The second equation in (\ref{eq:MC.series}) is satisfied if we take 
$$
\ep_2'=Q\psi_2=-Q\part_H(\ep_1\wedge\ep_1)' = \part_H Q(\ep_1\wedge\ep_1)'.
$$
Clearly it satisfies the condition 2'). 

Now suppose we have found the solutions for the first $n-1$ equations in (\ref{eq:MC.series}), $n>2$, such that $\bpart_H^\dagger\ep_i'=0$ and $\ep_i' = \part_H\varphi_i$ for all $2\le i<n$. We construct below a solution for the $n$-th equation:
\begin{equation}
\label{eq:n}
\bpart_H\ep_n' \;=\; - \frac12\sum_{k=1}^{n-1}[\ep_k,\ep_{n-k}]',
\end{equation}
which satisfies the condition 2'). Let $\psi_n$ denote the right-hand side of (\ref{eq:n}). By assumption, $\part_H\ep_i'=0$ for all $i<n$. Corollary \ref{cor:identity2} shows that $\psi_n$ is $\part_H$-exact, so $\HH\psi_n=0$. In addition, one has
\begin{eqnarray}
-2\bpart_H\psi_n
        &=& \sum_{k=1}^{n-1}\left(d_E[\ep_k,\ep_{n-k}]\right)'\no\\
        &=& \sum_{k=1}^{n-1}\left([d_E\ep_k,\ep_{n-k}]
                -[\ep_k,d_E\ep_{n-k}]\right)'\no\\
        &=& -\frac12\sum_{k=1}^{n-1}\sum_{i=1}^{k-1}
                \left([[\ep_i,\ep_{k-i}],\ep_{n-k}]\right)'
                +\frac12\sum_{k=1}^{n-1}\sum_{i=1}^{n-k-1}
                \left([\ep_k,[\ep_i,\ep_{n-k-i}]]\right)'\no\\
        &=& 2\sum_{\overset{1\le i<j<k}{i+j+k=n}}\Big(\big[\ep_i,[\ep_j,\ep_k]\big]
                +\big[\ep_j,[\ep_k,\ep_i]\big]+\big[\ep_k,[\ep_i,\ep_j]\big]\Big)'\no\\
        &&      + \sum_{\overset{1\le i\neq k}{2i+k=n}}\Big(2\big[\ep_i,[\ep_i,\ep_k]\big]
                +\big[\ep_k,[\ep_i,\ep_i]\big]\Big)'\no\\
        &&      + \sum_{\overset{1\le i}{3i=n}}\big[\ep_i,[\ep_i,\ep_i]\big]'\no\\
        &=& 0\no
\end{eqnarray}
 where in the last step we used the Jacobi identity for the Schouten bracket. Just as in the case of $\psi_2$, $\ep_n' = Q\psi_n$ solves (\ref{eq:n}), with condition 2') satisfied.
 
Convergence properties of $\ep(t)$ can be established by using the standard argument in deformation theory. For the sake of completeness, we give a proof that follows closely Ref.~\cite{KNS}. The only difference in our proof is that we work with a different norm on the space of differential forms than the one used in \cite{KNS}. 

Using the Hermitian metric introduced earlier, one can define the Sobolev $k$-norm on the space of complex differential forms in the usual way. Note that the choice of the Hermitian metric does not affect the equivalence class of the Sobolev $k$-norm. For a differential form $\al$, let us denote its Sobolev $k$-norm  by $||\al||_k$. Let $W^k$ be the space of sections of $\wedge^\bul T_\CC^*$ with bounded Sobolev $k$-norm, and one has the following filtration of spaces
$$
\Om^\bul(M,\CC)\equiv C^\infty(\wedge^\bul T_\CC^*)=W^\infty\subset\cdots\subset W^k\subset\cdots \subset W^1\subset W^0=L^2(\wedge^\bul T_\CC^*).
$$
For elementary properties of Sobolev spaces that is suited for our discussion, see for example refs.~\cite{GH,Wells}. The proof of the convergence of $\ep'(t)$ relies on the following lemma:
\begin{lem} \label{lem:inequalities}
Let $k$ be a positive integer. There exist positive constants $c_1$ and $c_2$ such that $\forall\,\phi\in C^\infty(\wedge^\bul E^*)$,
\begin{eqnarray}
||Q\phi'||_k &\le& c_1||\phi'||_{k-1},\no\\
||\part_H\phi'||_{k-1} &\le& c_2||\phi'||_k.\no
\end{eqnarray}
If $k>2N$, where $N=\dim_\CC M$, there exists a constant $c_3>0$ such that $\forall\,\phi,\psi\in C^\infty(\wedge^\bul E^*)$,
\begin{equation}
||(\phi\wedge\psi)'||_k \le c_3 ||\phi'||_k\cdot||\psi'||_k.\nonumber
\end{equation}
Furthermore, $c_1,c_2$ and $c_3$ depend only on $k$ and the manifold $M$.
\end{lem}
\begin{prf}
The first two inequalities are standard results on elliptic operators in Sobolev spaces, and the third one is a direct consequence of the fact that if $k>\dim_\RR M$, then there exists a positive constant $c$, which only depends on $k$ and $M$, such that $||fg||_k \le c||f||_k ||g||_k$ for any $f,g\in C^\infty(M)$.
\end{prf}

Let $k>2\dim_\CC M$ and choose $c_1, c_2$ and $c_3$ as in Lemma \ref{lem:inequalities}. Following \cite{KNS}, let us introduce the following power series
$$
a(t) = \frac{b}\la\sum_{n=1}^\infty \frac{\la^n t^n}{n^2}
$$
where $b$ is a positive constant such that $b\ge ||\ep_1||_k$, and $\la = 8c_1c_2c_3b$. It is shown in \cite{KNS} that\footnote{Given two series $a(t)=\sum_n a_n t^n$ and $b(t)=\sum_n b_n t^n$, the symbol $a(t)<<b(t)$ means  $b(t)$ is a majorant of $a(t)$, i.e., $b_n\ge |a_n|$ for all $n$.}
$$
a(t)^2 << 16\frac{b}\la a(t) = \frac2{c_1c_2c_3}a(t).
$$
We show that $\ep'(t)<< a(t)$ with respect to the Sobolev $k$-norm. For $n=1$, this is obvious by the above choice of $b$ and $\la$. Now assume $||\ep_i'||_k\le a_i$ for all $i\le n-1$. For $i=n$, we have
\begin{eqnarray}
||\ep_n'||_k &=& ||Q\psi_n||_k\no\\
&\le&c_1||\psi_n||_{k-1}\no\\
&\le&\frac12 c_1\sum_{i=1}^{n-1} ||\part_H(\ep_i\wedge\ep_{n-i})'||_{k-1}\no\\
&\le& \frac12 c_1c_2c_3\sum_{i=1}^{n-1}||\ep_i'||_k||\ep_{n-i}'||_k\no\\
&\le& \frac12 c_1c_2c_3 \left[a(t)^2\right]_n\no\\
&\le& a_n\no
\end{eqnarray}
This shows $\ep'(t)<<a(t)$ in Sobolev $k$-norm. As $a(t)$ is convergent in $|t|<1/\lambda$, $\ep'(t)\in W^k$. Note that $k$ can always be chosen such that $k=\dim_\CC M + m +1$ for some positive integer $m\ge2$. By Sobolev's fundamental lemma, $W^k\subset C^m(U_2)$, so we have $\ep'(t)\in C^m(U_2)$ or, equivalently, $\ep(t)\in C^m(\wedge^2 E^*)$. Acting on the Maurer-Cartan equation by $d_E^\dagger$, and noting that $d_E^\dagger\ep(t)=0$ by construction, we obtain the following quasi-linear PDE on $\ep$:
$$
\Delta_{d_E} \ep+\frac12 d_E^\dagger[\ep,\ep]=0.
$$
By a general result of PDE theory, $\ep(t)$ is smooth, i.e., $\ep(t)\in C^\infty(\wedge^2E^*)$. 
\end{prf}

Theorem \ref{thm:vanishing} tells us that, for a compact generalized Calabi-Yau manifold $(M,\J;\I)$, any infinitesimal deformation of $\J$ can be integrated into a finite one. Given such a finite deformation of $\J$, we have the following result on the deformation of the associated weak generalized Calabi-Yau structure $(M,\J)$. 

\begin{prp}
\label{prp:stability}
Let $\ep\in C^\infty(\wedge^2E^*)$ be a solution to the Maurer-Cartan equation (\ref{eq:MC}), so that it denotes a finite deformation of $\J$. If $\ep$ is sufficiently small, the deformed generalized complex structure $\J_\ep$ still defines a weak generalized Calabi-Yau structure, with the deformed generalized holomorphic volume form given by $\Om_\ep = \exp(-\ep)\cdot\Om$.
\end{prp}
\begin{prf}
We need to show that $\Om_\ep$ is a nowhere zero $d_H$-closed section of the canonical bundle for the deformed generalized complex structure $\J_\ep$. The $+i$-eigenbundle of $\J_\ep$ is spanned by elements of the form $A_\ep=(1+\ep)A \equiv A+A\lrcorner\ep\in C^\infty(E_\ep)$, where $A\in C^\infty(E)$. It is easy to show that
$$
A_\ep\cdot\rho = e^{-\ep}\cdot A\cdot e^{\ep}\cdot\rho, \qquad \forall \,\rho\in\Om^\bul(M,\CC).
$$
This immediately shows that $A_\ep$ annihilates $\Om_\ep$, so that $\Om_\ep$ is a section of the canonical bundle of $\J_\ep$. For $\ep$ sufficiently small, $\Om_\ep$ is also nowhere zero.

It then remains to show that $d_H\Om_\ep=0$. Using Lemma \ref{lem:identity} and Corollary \ref{cor:identity2} , one has
\begin{eqnarray}
\bpart_H(\ep^n\cdot\Om) &=& -\frac{n}2\,\ep^{n-1}\cdot[\ep,\ep]'\no\\
\part_H(\ep^{n+1}\cdot\Om) &=&  -\frac12 n(n+1)\,\ep^{n-1}\cdot[\ep,\ep]'\no
\end{eqnarray}
It follows immediately that
$$
d_H\Om_\ep = (\part_H+\bpart_H)\sum_{n\ge1}\frac{(-1)^n}{n!}\ep^n\cdot\Om = 0.
$$
\end{prf}

\section{Beyond the Classical Deformations}
\label{sec:extended}

The deformations considered so far are associated with actual variations of the generalized complex structure. There is in a sense a more general type of deformations, if one takes a more algebraic point of view. Recall that a generalized complex structure $\J$ is naturally associated with a differential Gerstenhaber algebra\footnote{We remind that a differential Gerstenhaber algebra is a differential graded algebra $(A,\wedge,d)$ equipped with a grade-$(-1)$ bracket $[\;\,]$, such that the latter is compatible with the graded commutative associative product $\wedge$, and that $d$ is a graded derivation of $[,]$.}
\begin{equation}
\label{eq:dga}
\g=\bigoplus_{k=1}^{2n}C^\infty(\wedge^k E^*),
\end{equation}
where the differential is the Lie algebroid differential $d_E$, and the bracket is the Schouten bracket on $\wedge^\bul E^*$. Instead of limiting ourselves to actual variations of $\J$, one can consider deformations of the entire algebraic structure of $\g$ as a differential Gerstenhaber algebra. 

Modern deformation theory addresses this question by assigning to $\g$ a deformation functor $\Def_\g$, which on the general level associates an Artin algebra to a set. We shall not go into details of the abstract deformation theory here, which is discussed amply in the literature (see, for example, refs.~\cite{GM,SS,Kon}). However we note the important fact that if $\g$ and $\g'$ are quasi-isomorphic, then $\Def_\g\simeq\Def_{\g'}$. In particular, if $\g$ is quasi-isomorphic to an abelian graded algebra $\h$\footnote{An abelian graded algebra can be regarded as a differential Gerstenhaber algebra with trivial differential and bracket.}, then the functor $\Def_\g$ is isomorphic to $\Def_\h$. The latter can be shown to be equivalent to the Hom-functor of the algebra of functions on a formal space $\M_\g$. In such a case, $\Def_\g$ is said to be representable, and $\M_\g$ is called the formal moduli space of $\g$. 

If the differential Gerstenhaber algebra defined in (\ref{eq:dga}) comes from a compact (twisted) generalized Calabi-Yau manifold $(M,\J;\I)$, one can show (as we shall do presently) that $\Def_\g$ is representable and the associated formal moduli space is a true, smooth manifold, which we denote by $\hM$. We refer to $\hM$ as the extended moduli space of $\J$. In addition to being a smooth manifold, it possesses the structure of a Frobenius manifold. In the special case that $\J$ comes from an ordinary complex structure, the extended moduli space is the one constructed and analyzed by Barannikov and Kontsevich in Ref.~\cite{BaKon}. Merkulov subsequently showed that the construction of \cite{BaKon} could be applied also to symplectic manifolds that satisfy the hard Lefschetz condition \cite{Merk}. In a broad sense\footnote{As in Sec.~\ref{sec:GCY}, the assumption of generalized Calabi-Yau structure (as in Definition \ref{def:GCY}) is not necessary; all the results in this section hold for (twisted) weak generalized Calabi-Yau manifolds that satisfy the generalized $\part\bpart$-lemma. In particular, this includes the scenario considered in Ref.~\cite{Merk}, as symplectic manifolds satisfying the hard Lefschetz condition satisfy a version of the $\part\bpart$-lemma.}, our results include those of refs.~\cite{BaKon,Merk} as special examples.

The moduli space $\M$ of actual deformations of $\J$, which we studied in detail in Sec.~\ref{sec:GCY}, is included as a subspace in $\hM$. We call deformations living in $\M$ the classical deformations, and those living in $\hM\backslash \M$ the non-classical deformations. Although the geometric meaning of the non-classical deformations is somewhat obscure, their physical interpretation is quite natural, as we shall discuss in Sec.~\ref{sec:physics}. 

\subsection{The extended moduli space for a generalized Calabi-Yau manifold}

Let $(M,\J;I)$ be a compact $H$-twisted generalized Calabi-Yau manifold, with $\Om$ being a generalized holomorphic volume form. Let $E$ be the $+i$-eigenbundle of $\J$. As before, we denote the generalized Dolbeault operators associated with $\J$ by $\part_H$ and $\bpart_H$. Let $\g$ be the differential Gerstenhaber algebra defined in (\ref{eq:dga}). As in \cite{BaKon}, one introduces a degree-$(-1)$ differential $\Delta$ (not to be confused with the Laplacians introduced before), which is defined by
$$
(\Delta\al)\cdot\Om = \part_H(\al\cdot\Om), \qquad \forall\,\al\in C^\infty(\wedge^\bul E^*).
$$
As a direct consequence of Lemma \ref{lem:identity}, the operator $\Delta$ satisfies the following algebraic relation: for any homogeneous $A, B\in\g$, 
\begin{equation}
\label{eq:BV}
(-1)^{|A|-1}[A,B] = \Delta(A\wedge B)-\Delta A\wedge B-(-1)^{|A|} A\wedge\Delta B.
\end{equation}
Namely, the Schouten bracket measures of the failure of $\Delta$ being a graded derivation for the associative produce $\wedge$. Here we denote the grade of an element $\al$ by $|\al|$. Equipped with $\Delta$, $\g$ acquires the additional structure of a BV algebra.

The associated formal moduli space can be constructed following the general lines of Ref.~\cite{BaKon}. Let $\h=H^\bul(\Delta)$, which is isomorphic to $H^\bul(d_E)$. We endow it with the structure of a graded vector space by assigning degree $p-2$ to elements in $H^p(\Delta)$. By a standard argument, it follows from Eq.~(\ref{eq:BV}) and the $\part_H\bpart_H$-lemma that $\g$ and $\h[-2]$ are quasi-isomorphic differential Gerstenhaber algebras; thus, $\Def_\g$ is representable. More specifically, if we define $\th_a$ to be a homogeneous basis of $\h$, and $t^a$ a set of dual basis with degree $|t^a|=-|\th_a|$, then $\Def_\g$ is represented by $\CC[[t_\h]]$, the algebra of formal power series on $\h$. It follows immediately that the formal moduli space $\hM$ associated to $\g$ is a smooth formal manifold of complex dimension $\sum_k \dim H^k(d_E)$.

For many physical and geometrical applications, knowing the existence of $\M$ as a formal manifold is usually not good enough; instead, one often wants to know whether the moduli space exists as a smooth manifold in the usual sense. For example, this is the case when one studies admissible deformations of the generalized B-model defined on a generalized Calabi-Yau manifold, as we shall discuss in Sec.~\ref{sec:physics}. A formal deformation represented as a power series in $t_\h$ is physically acceptable if and only if it converges uniformly in a small neighborhood of the origin. 

To address this question, let us go back to the definition of the formal moduli space $\hM$: it is the set of solutions to the extended version of the Maurer-Cartan equation
\begin{equation}
\label{eq:MC_e}
d_E\hep + \frac12[\hep,\hep] = 0, \qquad \hep\in \hg^2,
\end{equation}
 modulo gauge equivalence. Here $\hg$ is the graded tensor product $\g\hotimes\CC[[t_\h]]$. The differential $d_E$ and the Schouten bracket naturally extend to $\hg$. The question then boils down to whether one can construct a uniformly convergent solution to the extended Maurer-Cartan equation for every element in $H^\bul(d_E)$. This turns out to be true, as one can check by repeating the analysis almost exactly as in the proof of Theorem \ref{thm:vanishing}. We simply state the result.
 
\begin{thm}
\label{thm:GCY_e}
There exists a formal power series solution to the extended Maurer-Cartan equation (\ref{eq:MC_e}) in the form of
$$
\hep(t) = \sum_a \hep_a t^a + \frac12\sum_{a_1,a_2}\hep_{a_1a_2}t^{a_1}t^{a^2} + \cdots
$$
such that
\begin{itemize}
\item{$\hep_a'\equiv \hep_a\cdot\Om$ form a basis of the harmonic forms;}
\item{All higher order terms $\hep_{a_1\ldots a_k}$ for $k\ge2$ are $\Delta$-exact and $d_E^\dagger$-closed;}
\item{The power series $\hep(t)$ is uniformly convergent in a neighborhood of $t=0$.}
\end{itemize}
\end{thm}
The coordinates $t_\h$ are called the flat coordinates on $\hM$, for reasons that will become apparent later.
 
\begin{cor}
The extended moduli space of $\J$ on a compact (twisted) generalized Calabi-Yau manifold $(M,\J;\I)$ is un-obstructed and smooth, and is of dimension $\sum_p \dim H^p(d_E)$. 
\end{cor}

\subsection{The Frobenius structure and flat coordinates}

In this subsection we show that the extended moduli space $\hM$ is a Frobenius manifold. In the  case of ordinary Calabi-Yau manifolds, this fact is established by Barannikov and Kontsevich \cite{BaKon} (see also the  discussion in Ref.~\cite{Manin}). For generalized Calabi-Yau manifolds, initial steps in this direction have already been taken in Ref.~\cite{KapLi}. 

Let us first recall how integration is defined as a linear functional on $\g$ in Ref.~\cite{KapLi}. Given a generalized holomorphic volume form $\Om$, one defines its conjugate $\tOm$ as follows. Let $\iota$ be the automorphism\footnote{This operation was also used by Gualtieri in Ref.~\cite{Gua1}, and had appeared before in the literature of Dirac structure. We thank M.~Gualtieri for pointing this out to us.} of $T\oplus T^*$ that maps $(X,\xi)$ to $(X,-\xi)$. It is easy to check that $\tilde\J\equiv \iota^{-1}\J \iota$ is again a generalized complex structure, but twisted by $\tilde{H}=-H$; in fact, $(M,\tilde\J)$ defines a weak generalized Calabi-Yau structure twisted by $\tilde{H}$. The conjugate $\tOm$ is then defined to be a generalized holomorphic volume form of $(M,\tilde\J)$\footnote{We note that $\tOm$ is {\em not} the complex conjugate of $\Om$. In physical terms, if $\Om$ is interpreted as a ground state in the Ramond-Ramond sector of the $(2,2)$ supersymmetric sigma model, then $\tOm$ is the analog of the so-called BPZ conjugate of $\Om$.}. One can fix the ambiguity by requiring $\tOm$ to have the same norm as $\Om$. Given $\al\in \g$, we define its integration to be
\begin{equation}
\label{eq:integral}
\int \al \equiv \int_M \tOm\wedge(\al\cdot\Om).
\end{equation}
\begin{prp}
\label{prp:identities}
Let $\al,\be$ be homogeneous elements in $\g$. Then the following identities hold:
\begin{eqnarray}
\int d_E\al\wedge\be &=& (-1)^{|\al|+1}\int \al\wedge d_E\be\nonumber\\
\int \al\wedge \Delta\be &=& (-1)^{|\al|}\int \Delta\al\wedge\be.\nonumber
\end{eqnarray}
\end{prp}
\begin{prf}
The first identity is essentially proven in Ref.~\cite{KapLi}, where it is shown that the integral (\ref{eq:integral}) for a $d_E$-exact element vanishes. The second identity holds since
\begin{eqnarray}
\int \al\wedge \Delta\be &=& \int_M \tOm\wedge \al\cdot \part_{\J,H}(\be\cdot\Om)\nonumber\\
&=& (-1)^{|\al|(|\tOm|+1)}\int_M (\tilde\al\cdot\tOm)\wedge \part_{\J,H}(\be\cdot\Om)\nonumber\\
&=& (-1)^{(|\al|+1)|\tOm|+1}\int_M \part_{\tilde\J,\tilde{H}}(\tilde\al\cdot\tOm)\wedge \be\cdot\Om\nonumber\\
&=&(-1)^{(|\al|+1)|\tOm|+1+(|\al|-1)(|\tOm|+1)}\int_M \tOm\wedge \Delta\al\wedge \be\cdot\Om\nonumber\\
&=& (-1)^{|\al|}\int \Delta\al\wedge\be.\nonumber
\end{eqnarray}
\end{prf}

Given a solution $\hep(t)$ of the Maurer-Cartan equation as stated in Theorem \ref{thm:GCY_e}, one can construct a differential on $\hg$ given by
$$
D_E \equiv d_E + [\hep,\cdot].
$$
Let $\part_a=\part^L/\part t^a$ be the left derivatives with respect to $t^a$, and $\cO_a\equiv \part_a\hep(t)$. It is easy to show that $\cO_a$ live in the cohomology of $D_E$; they form a local basis in the tangent space of $\M$ at $\hep$. The metric on $\M$ is then defined by
$$
G_{ab} = \int \cO_a\wedge \cO_b.
$$
From the form of the solution $\hep(t)$ given in Theorem \ref{thm:GCY_e}, $\cO_a= \hep_a + \Delta\rho_a$ for some $\rho_a$. It follows immediately that $G_{ab}(t)=G_{ab}(0)$; namely, the metric $G$ is locally constant in the flat coordinates $t$. Physically, this means that the two-point functions are invariant under deformations.

Another ingredient of the Frobenius structure is a tensor $C_{ab}^{\;\;c}$, defined by
\begin{equation}
\label{eq:opt}
\cO_a\wedge\cO_b = C_{ab}^{\;\;c}\cO_c + D_E\cO_{ab},
\end{equation}
where $\cO_{ab}$ is defined only up to $D_E$-closed terms. We use the locally constant metric $G$ to raise and lower the indices: $C_{abc}\equiv C_{ab}^{\;\;\,d}G_{dc}$. 

\begin{prp}
There exists a potential $S$ such that $C_{abc} = \part_a\part_b\part_c S$ in the flat coordinates.
\label{prp:potential}
\end{prp}
\begin{prf}
Let $\hep(t)=\sum_a \hep_a t^a + \Delta\phi\in\hat\g^2$ be a solution to the Maurer-Cartan equation (\ref{eq:MC_e}) that satisfies the conditions in Theorem \ref{thm:GCY_e}.  The potential is given by
$$
S = \frac12\int d_E\phi\wedge \Delta\phi + \frac16 \int \hep\wedge\hep\wedge\hep.
$$
The rest of the proof follows the calculation in the appendix of Ref.~\cite{BaKon}.
\end{prf}

One can also introduce the generalized periods as in \cite{BaKon}. Let $\hep(t)\in\hg^2$ be a solution to the Maurer-Cartan equation as given in Theorem \ref{thm:GCY_e}. We define the deformed generalized holomorphic volume form by
$$
\hat\Om(t) \equiv e^{-\hep(t)}\cdot\Om.
$$
By the same argument as in Proposition~\ref{prp:stability}, one easily shows that $\hat\Om(t)$ is $d_H$-closed. Let $\th^a$ be the dual basis of $\part_a$ on the tangent bundle of $\hM$. The generalized periods are then defined by $\Pi_i\equiv \Pi_{ia}\th^a$, where
$$
\Pi_{ia} = \int_{C_i} \part_a\hat\Om(t).
$$
In the above expression, $\{C_i\}$ denote a basis of the dual of the twisted de Rham cohomology $H^\bul(d_H)$, and the integration sign denotes evaluation of $[\part_a\hat\Om]\in H^\bul(d_H)$ on $C_i$. In the special case of $H=0$, $\Pi_{ia}$ are simply integration of $\part_a\hat\Om$ on homology cycles $C_i$ that represent a basis of $H_\bul(M,\CC)$. 

The structure constants $C_{ab}^{\;\;\;c}$ then define a natural connection $\nabla$ on $T\hM$ via
$$
\nabla_{\part_a}\th^b = C_{ac}^{\;\;\;b}\th^c.
$$
Just like in the more special case studied in Ref.~\cite{BaKon}, one can show that $\nabla$ is a flat connection. To this end, it suffices to assume that $\part_a$ is even. From the Maurer-Cartan equation, we have
$$
D_E\part_a\part_b\hep(t) = \Delta(\part_a\hep(t)\wedge\part_b\hep(t)) = \Delta D_E\cO_{ab},
$$
where $\cO_{ab}$ is defined in (\ref{eq:opt}). From the generalized $\part\bpart$-lemma for $D_E$ and $\Delta$ \cite{Manin}, we have $\part_a\part_b\hep+\Delta\cO_{ab}=\Delta D_E f_{ab}$ for some $f_{ab}$. As $\cO_{ab}$ is only defined up to $D_E$-closed terms, we can absorb $D_E f_{ab}$ into the definition of $\cO_{ab}$. Therefore, there exists $\cO_{ab}$ such that
$$
\part_a\hep\wedge\part_b\hep-\part_a\part_b\hep = A_{ab}^{\;\;\;c}\part_c\hep + (\Delta+ D_E)\cO_{ab}.
$$
It then follows that
\begin{eqnarray}
\part_a \Pi_i &=& \int \left(\part_a\hep\wedge\part_b\hep-\part_a\part_b\hep\right)\cdot\hat\Om(t)\cdot\th^b\no\\
&=& -A_{ab}^{\;\;\;c}\Pi_{ic}\th^b.\no
\end{eqnarray}
In the last line above, we have used the relation
$$
(\Delta\cO+D_E\cO)\cdot \hat\Om(t) = d_H(\cO\cdot\hat\Om(t)),
$$
a fact that one can check easily. This immediately shows that $\nabla_{\part_a}\Pi_i=0$ for all $i$. Since $\Pi_i$ form a local frame on $T\hM$, we have:
\begin{cor}
The connection $\nabla$ is flat.
\end{cor}

\section{Physical Interpretations}
\label{sec:physics}

The unobstructedness of the extended moduli space $\hM$ has direct physical implications, since $\hM$ parameterizes admissible deformations of the generalized B-model. To explain this connection, we need some physical background. The generalized B-model \cite{KapLi} is a topological sigma model defined on a (twisted) generalized Calabi-Yau  manifold $M$. It is equipped with a nilpotent BRST operator $Q_B$, and is an example of cohomological field theories. One way to construct the generalized B-model is to regard it as a `twisted version' of the $(2,2)$ supersymmetric sigma model on $M$. At the quantum level, the generalized B-model is a well-defined cohomological field theory if and only if the target manifold admits a (twisted) generalized Calabi-Yau structure $(M,\J;\I)$ \cite{KapLi}. Furthermore, the BRST operator $Q_B$ can be identified with $d_E$, the Lie algebroid differential associated with the $+i$-eigenbundle of $\J$, and the space of observables is identified with the Lie algebroid cohomology $\oplus_p H^p(d_E)$. In the special case when $M$ is an ordinary Calabi-Yau manifold, the generalized B-model reduces to the more familiar A-model and B-model discovered by Witten \cite{Witten.Mirr}.
 
The space of observables parameterizes possible deformations of the theory. In the case of a cohomological field theory, deformations can be explicitly constructed as perturbations to the original Lagrangian by using the descent equations. Even though such a deformation is automatically annihilated by the original BRST operator, it can only be regarded as an infinitesimal deformation of the generalized B-model as a cohomological field theory. The reason is that after the perturbation, there is no guarantee that the new theory still admits a nilpotent BRST operator. The question whether additional admissible perturbations can be added to the Lagrangian to restore the nilpotency of the BRST operator is nontrivial in general. For ordinary B-model defined on Calabi-Yau manifolds, this analysis was initiated by Witten \cite{Witten.Mirr} and later completed by Barannikov and Kontsevich \cite{BaKon}.

The results in secs.~\ref{sec:GCY} and \ref{sec:extended} show that for the generalized B-model defined on a twisted generalized Calabi-Yau manifold, an infinitesimal deformation can always be completed into a finite deformation such that the perturbed theory is still cohomological. In this context, the extended moduli space $\hM$ is nothing but the moduli space for the family of cohomological field theories, with the base point being the original unperturbed generalized B-model. Other results of Sec.~\ref{sec:extended} also  translate easily to this physical setup: the BRST operator of the deformed theory is identified with $D_E$; an independent basis of observables in the deformed theory can be taken to be $\cO_a=\part_a\hep$; the topological metric on the space of observables is given by the two-point functions $G_{ab}$, and from Sec.~\ref{sec:extended} we know that it receives no corrections in the extended moduli space; the three-point functions are given by the tensor $C_{abc}$ that comes from a single potential $S$ as given in Proposition~\ref{prp:potential}. In the special case of $\dim_\RR M=6$ that is of interest in string theory applications, the potential $S$ can be interpreted as the classical action of the string field theory of the generalized B-model.

The attentive reader might have noticed a subtlety in the above reasoning. Consider a small (but finite) deformation living in the {\em classical} moduli space $\M$. By definition, it corresponds geometrically to an integrable deformation of the generalized complex structure $\J$ and is parameterized by some $\ep\in C^\infty(\wedge^2E^*)$ that satisfies the Maurer-Cartan equation. The deformed theory is again a cohomological field theory, but the relevant underlying generalized complex structure is now $\J_\ep$. From this perspective, one expects the BRST operator of the deformed theory to be given by the Lie algebroid differential associated with $\J_\ep$, namely $d_{E_\ep}$. On the other hand, we have identified the deformed BRST operator with $D_E=d_E+[\ep,\bul]$ in the last paragraph. However, $d_{E_\ep}$ and $D_E$ are different operators. From the discussion in Sec.~\ref{sec:extended}, $D_E$ is a deformation of the original Lie algebroid differential $d_E$ and it preserves the grading of $\g$ defined in (\ref{eq:dga}); $d_{E_\ep}$, on the other hand, does not preserve the grading of $\g$.  

The way out of this apparent contradiction is to note that although $d_{E_\ep} \neq D_E$, the differential complexes associated with them are isomorphic. To see this, let $\sm: E\to E_\ep$ be the isomorphism given by 
$$ 
\sm: \; A\mapsto A_\ep \equiv A + A\lrcorner\ep, \quad \forall A\in C^\infty(E),
$$
and $\sm^*: E_\ep^*\to E^*$ be its dual. They can be promoted naturally to maps of complexes 
$$
\sm: \wedge^\bul E\to\wedge^\bul E_\ep, \quad 
\sm^*: \wedge^\bul E_\ep^* \to \wedge^\bul E^*.
$$
Then we have the following proposition, whose proof can be found in Appendix.

\begin{prp}
\label{prp:isomorphism}
The map $\sm^*$ is an isomorphism of complexes $(\g, D_E)$ and $(\g_\ep, d_{E_\ep})$, i.e.,
$$
d_{E_\ep} = {\sm^*}^{-1}\cdot D_E\cdot\sm^*.
$$
\end{prp}
Therefore, it is justified to identify $D_E$ as the deformed BRST operator.


\section*{Appendix}
In this appendix, we give a proof of Proposition \ref{prp:isomorphism}. We show that $\forall \,\om\in C^\infty(\wedge^\bul E^*)$,
\begin{equation}
\label{app.iso}
\sm^*\cdot d_{E_\ep}\cdot {\sm^*}^{-1} \om = D_E\,\om
\end{equation}
Let $\pi$ denote the projection from $T_\CC\op T^*_\CC$ to $T_\CC$. By restriction, it provides the anchor maps for all the Lie algebroids ($E, E^*, E_\ep$ and $E_\ep^*$) considered below. For $A\in C^\infty(E)$, we use the notations $A\lrcorner\ep$ and $\iota_A\ep$ interchangeably, and we denote $\sm(A)$ by $A_\ep$.

We proceed by induction on the degree of $\om$. First, consider the case $\om=f\in C^\infty(M)$. For any $A\in C^\infty(E)$, we have
\begin{eqnarray}
\sm^* d_{E_\ep} {\sm^*}^{-1}f (A) = d_{E_\ep}f (A_\ep) &=& \pi(A_\ep)f\no\\
&=& \pi(A)f + \pi(A\lrcorner\ep)f\no\\
&=& d_Ef(A) + [\ep, f](A) = D_Ef(A).\no
\end{eqnarray}
Therefore (\ref{app.iso}) holds for $\om\in C^\infty(M)$.

Next, assume $\om\in C^\infty(E^*)$.  For any $A, B\in C^\infty(E)$, we have
\begin{eqnarray}
\sm^*d_{E_\ep}{\sm^*}^{-1} \om (A, B) &=& 
        d_{E_\ep}({\sm^*}^{-1}\om) (A_\ep, B_\ep) \no\\
        &=& \pi(A_\ep){\sm^*}^{-1}\om   (B_\ep) - \pi(B_\ep){\sm^*}^{-1}\om     (A_\ep) -       {\sm^*}^{-1}\om ([A_\ep, B_\ep])\no\\
        &=& \pi(A_\ep)\om(B) - \pi(B_\ep)\om(A) -       \om(\sm^{-1}[A_\ep, B_\ep]).
\label{eq:app}
\end{eqnarray}
Using the definition of the Courant bracket, and that both $E$ and $E_\ep$ define generalized complex structures, it is easy to show that
$$
\sm^{-1}[A_\ep, B_\ep] = [A, B] + d_{E^*}\,\ep(A,B) + (\iota_A\ep)\lrcorner d_{E^*}B
                                                                - (\iota_B\ep)\lrcorner d_{E^*}A.
$$
Here $d_{E^*}$ denotes the Lie algebroid differential associated with $E^*$. Substituting it into (\ref{eq:app}), and after some algebra one obtains
\begin{eqnarray}
\sm^*d_{E_\ep} {\sm^*}^{-1} \om (A, B) &=& d_E\om(A,B)
                        + \pi(\om)\ep(A,B) + B\lrcorner[A\lrcorner\ep,\om] 
                        - A\lrcorner[B\lrcorner\ep,\om]\no\\
        &=& d_E\om(A,B)+ [\ep, \om](A,B)\no\\
        &=& D_E\,\om(A,B),\no
\end{eqnarray}
so (\ref{app.iso}) also holds for $\om\in C^\infty(E^*)$. Since both $D_E$ and $d_{E_\ep}$ are derivations, one can easily show that (\ref{app.iso}) holds for $\om$ with arbitrary degree by induction. This completes the proof of Proposition \ref{prp:isomorphism}.

\section*{Acknowledgments}
I am indebted to Marco Gualtieri and Anton Kapustin for correspondence and helpful discussions. I am also grateful to Ke Zhu for explaining aspects of Sobolev spaces to me. Part of this work was completed while I was visiting the Institute for Advanced Study and Rutgers University in May, 2005. This work was supported in part by the DOE
grant DE-FG02-92-ER40701.

\end{document}